# Predicting Pneumonia and Region Detection from X-Ray Images using Deep Neural Network

Sheikh Md Hanif Hossain[1], S M Raju[2], Amelia Ritahani Ismail[3]
Department of Computer Science
International Islamic University Malaysia
[1]sheikhhanifhossain@gmail.com, [2]rajuiium121@gmail.com, [3]amelia@iium.edu.my

*Abstract* – Biomedical images are increasing drastically. Along the way, many machine learning algorithms have been proposed to predict and identify various kinds of diseases. One such disease is Pneumonia which is an infection caused by both bacteria and viruses through the inflammation of a person's lung air sacs. In this paper, an algorithm was proposed that receives x-ray images as input and verifies whether this patient is infected by Pneumonia as well as specific region of the lungs that the inflammation has occurred at. The algorithm is based on the transfer learning mechanism where pre-trained ResNet-50 (Convolutional Neural Network) was used followed by some custom layer for making the prediction. The model has achieved an accuracy of 90.6 percent which confirms that the model is effective and can be implemented for the detection of Pneumonia in patients. Furthermore, a class activation map is used for the detection of the infected region in the lungs. Also, PneuNet was developed so that users can access more easily and use the services.

## I. Introduction

The medical health sector is one of the most important sectors in human society which saves thousands of human lives every day. There is immense pressure on doctors to check and treat every patient available. Humans are prone to error. It is only normal that doctors will make mistakes. But in medical sector, when doctors make mistakes, it is deadly and often results in unwanted death. Using the PneuNet, doctors can check the x-ray to make sure the verdict they reached was accurate or not. This system will help to reduce the pressure off from the doctors. Other than the doctors, patients can also check by themselves by uploading the x-ray picture in PneuNet to see whether he or she has pneumonia or not. This will reduce the cost of going to see the doctors again and again. For developing and under-developed countries, health care is expensive and often people cannot afford the treatment. This system can largely help the patients in these regions to check for pneumonia rather easily than going to the doctor's multiple times for checkup. This system will help to reduce the treatment cost for pneumonia which will help insolvent people to get proper treatment with less cost. For the health sector itself, this PneuNet will help to automate the process which will save time and resources. The doctors can treat more patients at a lower cost, and everyone is getting benefitted using this system.

## II. Previous Works

There has been a lot of works done in medical disease detection along with machine learning. However, for pneumonia detection, there has not been a lot of work done, only a handful of research was found regarding Pneumonia detection. Among them, some of the researchers used deep learning to detect pneumonia using image x-ray as well as generating heatmap to see the regions where the pneumonia was detected [1]. Other researchers worked on transfer learning to detect whether the person has pneumonia or not [2]. It was found out by comparing these two articles that transfer learning yielded better results than other algorithms. In some research, researcher used focal loss to prevent bias in the imbalanced dataset [3]. Many other researchers used Class Activation Map (CAM) to get the discriminative image regions used by a Convolutional Neural Network (CNN) to identify a specific class in the image [4]. ResNet-50 a Convolutional Neural Network which was trained on millions of images and used to classify thousand classes of images accurately [5].

From the previous works, it was found out that the ResNet-50 pre-trained model would be a good option to use for pneumonia detection. Along with ResNet-50, focal loss would be applied to increase the accuracy of imbalanced dataset which was used in this paper. Also, CAM was used to identify the specific region of the lungs that is inflamed by pneumonia. This CAM simply takes the final convolutional feature map and weighs every channel in that feature with the gradient of the loss with respect to the channel.

All the approach mentioned *above* was implemented in the PneuNet system. A web application was developed from the results of the PneuNet system to ease the use of users.

## III. Methodology

In this section, the process of developing PneuNet is discussed in brief.

### 3.1 Data Collection

The data was collected from one of the previous researches which was conducted on medical disease detection [6]. The data was open-source data.

### 3.2 Data Preprocessing and Visualization

The dataset was divided into train and test data. The dataset was organized into three folders, train, test, validation as well as contained subfolder for each image category, pneumonia and normal. There were 5863 x-ray images in jpeg format. Further visualization was done to see how many cases of pneumonia and normal x-ray images there in the training dataset were. The total number of training records was 5216 and total number of pneumonia cases was 3875 and the rest were normal cases. The number of test records was 634. Also, some visualizations were done to see the x-ray images of normal cases and pneumonia cases. Data augmentation was performed to increase the diversity of the training data without inserting new data into the training dataset. In the data augmentation, image flipping, and rotation were done to increase the number of training records. It will help to identify pneumonia cases much better and better train the model. Further works were done in the model building section.

### 3.3 Model

In this paper, pre-trained ResNet-50 was used as a base model. Image size of 224,224,3 was used as input for base model. The base model layers were frozen and followed by some custom layers were added to create the final architecture. The custom layers consist of a dense layer followed by a dropout layer and again a dense layer which was the output layer as well. The first dense layer had 50 neurons and for the activation function Rectified Linear Unit (ReLU) was used. Then in the dropout layer, dropout 0.5 was applied to all the neurons. In the final layer, sigmoid activation functions were used to get the final output of the model. The model was compiled using focal loss as a loss function and Root Mean Square Propagation (RMSProp) as the optimizer function. Furthermore, accuracy was used as the metric of the model.

While training, the batch sizes were 16 and epochs were 100. Early stopping mechanism was used to prevent overfitting. Early stopping mechanism is used when the model can no longer learn and start to overfit. CAM was used during the prediction process of pneumonia. The test set was used to evaluate the model and the results will be discussed in the *next* Results section.

### 3.4 PneuNet

The model was saved and deployed as a web application where the user will upload the image of x-ray and the model will identify and predict whether the uploaded x-ray image has pneumonia or not. And if the patient has pneumonia then the system will identify the region and show another image to the patient highlighted the infected region in his x-ray image.

## IV. Results

The final model has achieved an accuracy of 90.06 percent using the test dataset for predicting whether a patient has been infected by pneumonia. Furthermore, the precision and recall of our model are 92 and 93

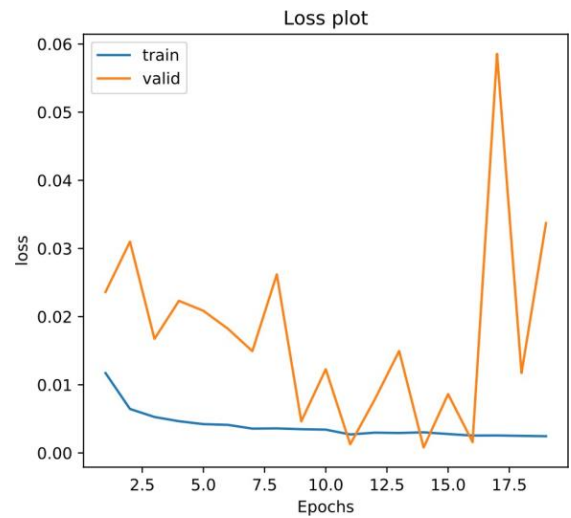

**Figure 1: Train vs validation loss**

percent, respectively. From Figure 1, it can be seen how the train and validation loss is decreasing over the number of epochs. Also, from figure 2, accuracy is increasing over the number of epochs for both training and validation. The confusion matrix was shown in

Figure 3. In Figure 4, the Receiver Operating

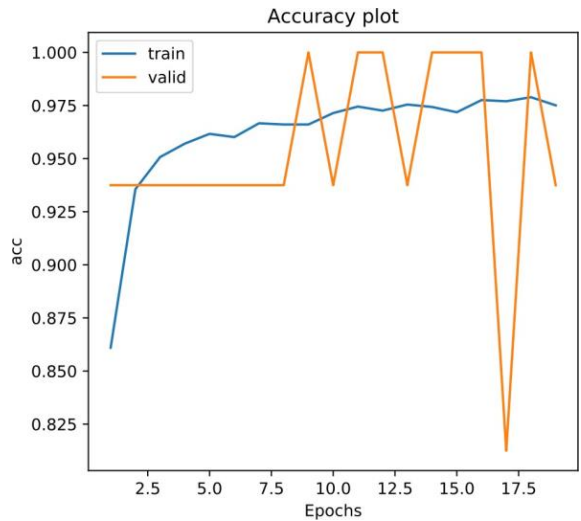

**Figure 2: Train vs validation accuracy**

Characteristic (ROC) curve was shown and the Area Under Curve (AUC) was 89 percent.

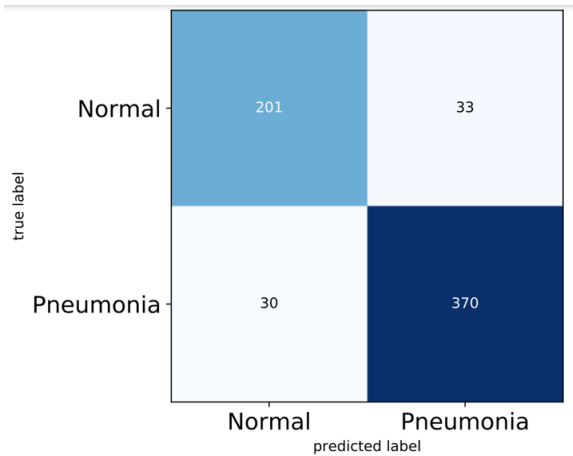

**Figure 3: Confusion matrix**

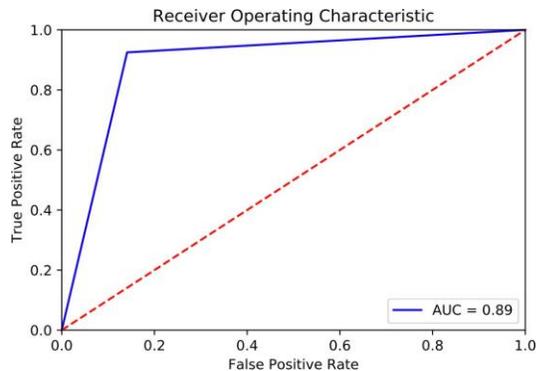

**Figure 4: ROC Curve**

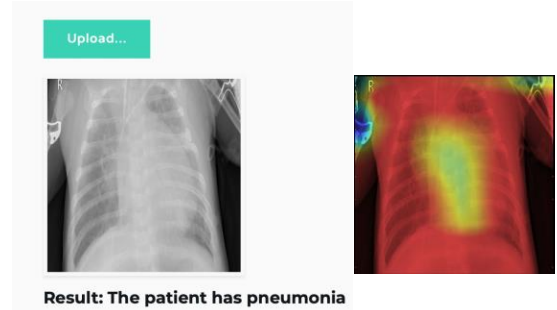

**Figure 5: PneuNet prediction**

Figure 5 shows that the patient who uploaded the x-ray images has pneumonia and the infected area is highlighted using the PneuNet system.

## V. Discussion

From the result section, the accuracy of this model is showing around 90 percent. This is a huge improvement over previous researches as their accuracy was around 77 percent [1]. Also, the precision and recall are around 92 and 93 percent, respectively. This shows that the performance of this model is quite good. For pneumonia, the cost of false prediction is very high. So, 92 percent precision and 93 percent recall mean the model can correctly identify and label the pneumonia patient correctly. The F1 score for this model is 0.92. Also, the AUC of this model is around 89 percent, that means the model is able to separate the classes of whether the patient has pneumonia or not pneumonia.

Focal loss helped to achieve high accuracy in this model with imbalanced dataset. Also, CAM helped to make the region detection from x-ray images. The dataset records were only around 5000. More dataset will aid in getting much higher accuracy.

Additionally, the results of this research were saved, and a product was built for the ease of user. None other research prior to that focused on the end product which will help the mass public.

## VI. Conclusion

According to World Health Organization (WHO) pneumonia accounts for 15% of all deaths of children under 5 years old, killing 808694 children in 2017. This number is increasing drastically each year especially in the third world countries due to lack of experts. Early diagnosis is the key to prevent such huge number of children's death. Chest X-rays are the most widely used method to diagnose pneumonia by medical experts. Due to shortage of experts and costly procedures for the third world countries pneumonia remains as a great threat of children's lives.

Our proposed PneuNet system can be widely used by public or radiologists for early detection of pneumonia. Confusion matrix of our model ensures that PneuNet is reliable thus by using the system, medical expenditure can be reduced in a large extend as well.